\documentclass[aps,showpacs,preprint]{revtex4}
\usepackage{graphicx}
\usepackage{epstopdf}
\usepackage{dcolumn}
\usepackage{amsmath}

\begin{document}
\title{A geometry for the shell model}
\author{P.~Van~Isacker}
\affiliation{Grand Acc\'el\'erateur National d'Ions Lourds (GANIL),
CEA/DRF--CNRS/IN2P3, Bvd Henri Becquerel, B.P.~55027, F-14076 Caen, France}

\date{\today}

\begin{abstract}
A geometric interpretation is given of matrix elements of a short-range interaction
between states that are written in terms of aligned neutron--proton pairs.
\end{abstract}

\maketitle

\section{Introduction}
\label{s_intro}
In this contribution matrix elements of the nucleon--nucleon interaction
between shell-model states are studied from a geometric point of view.
An approach of this kind goes back to the seminal study of Schiffer~\cite{Schiffer71}.
The interaction between two nucleons can be expressed
in terms of the angle between the angular momenta of the nucleons.
Furthermore, the strength of the nucleon--nucleon interaction
has a universal dependence on this angle---universality that can be shown to result
from the short-range character of the effective interaction~\cite{Molinari75}.
This geometric interpretation of the nucleon--nucleon interaction
is by now an accepted feature of nuclear structure;
a discussion of it can be found in textbooks on nuclear physics,
see e.g.\ Sect.~4.2 of Ref.~\cite{Casten00}.

The geometry of the shell model can be extended
from two- to four-nucleon configurations~\cite{Isacker14}.
In this contribution the geometric interpretation of the neutron--proton (np) interaction
is examined for a four-nucleon configuration
when written in terms of aligned np pairs.

\section{Matrix elements of the neutron-proton interaction in a 2n--2p basis}
\label{s_2n2p}
States for two neutrons with angular momentum $J_\nu$
and two protons with angular momentum $J_\pi$,
coupled to total angular momentum $J$,
can be written as
$|J_\nu J_\pi;J\rangle\equiv|j_\nu j_\nu(J_\nu),j_\pi j_\pi(J_\pi);J\rangle$,
which, for all even values of $J_\nu$ and $J_\pi$ allowed by angular-momentum coupling,
form a complete and orthogonal basis.
Matrix elements of the np interaction in the 2n--2p basis are
\begin{equation}
\langle J'_\nu J'_\pi;J|\hat V_{\nu\pi}|J''_\nu J''_\pi;J\rangle=
4[J'_\nu][J''_\nu][J'_\pi][J''_\pi]
\sum_R(2R+1)
\left[\begin{array}{cccccccc}
\!\!j_\nu\!\!&&\!\!j_\pi\!\!&&\!\!J'_\pi\!\!&&\!\!J'_\nu\!\!&\\
&\!R\!&&\!\!j_\pi\!\!&&\!\!J\!\!&&\!\!j_\nu\!\!\\
\!\!j_\nu\!\!&&\!\!j_\pi\!\!&&\!\!J''_\pi\!\!&&\!\!J''_\nu\!\!&
\end{array}\right]
V^R_{\nu\pi},
\label{e_me2n2p}
\end{equation}
with $[x]\equiv\sqrt{2x+1}$,
and where $V^R_{\nu\pi}\equiv\langle j_\nu j_\pi;R|\hat V_{\nu\pi}|j_\nu j_\pi;R\rangle$
are the np two-body matrix elements
and the symbol in square brackets is a $12j$ coefficient of the second kind,
which is known as a sum of products of four $6j$ coefficients~\cite{Yutsis62}.

The expression~(\ref{e_me2n2p}) is used in Ref.~\cite{Kimun}
to demonstrate the crucial role played by the np matrix element $V^R_{\nu\pi}$
with aligned neutron and proton angular momenta,
that is, for $R=j_\nu+j_\pi$.
This generic feature of the np interaction results
from angular momentum coupling
[i.e., from the $12j$ coefficient in Eq.~(\ref{e_me2n2p})]
as well as from the matrix element in the aligned configuration,
which is moderately to strongly attractive for any reasonable np interaction.
The importance of the aligned np matrix element
raises the question whether a shell-model approximation
can be formulated in terms of aligned np pairs,
as recently proposed in Ref.~\cite{Cederwall11}.
(See Sect.~4.3 of Ref.~\cite{Frauendorf14} for a review of this debate.)
This question is also addressed in Ref.~\cite{Kimun} for a 2n--2p system,
and more generally for $k$n--$k$p systems in $N=Z$ nuclei in Ref.~\cite{Isacker11},
with the conclusion that the wave functions of many, but certainly not all, yrast states 
have a dominant component in terms of aligned np pairs.
Here I examine whether a description in terms of aligned np pairs
gives rise to a geometry of the shell model.

\section{Matrix elements of the neutron-proton interaction in a np--np basis}
\label{s_npnp}
An alternative basis exists in terms of np pairs of the form
$|J_1J_2;J\rangle\!\rangle\equiv|j_\nu j_\pi(J_1),j_\nu j_\pi(J_2);J\rangle\!\rangle$.
If $J_1$ and $J_2$ acquire all values allowed by angular-momentum coupling,
this defines an overcomplete basis of states,
which are not necessarily normalized,
as indicated by the double bracket.
Matric elements in the np--np basis
can be obtained with the help of the transformation
\begin{equation}
|J_1J_2;J\rangle\!\rangle=
-\sum_{J_\nu\;{\rm even}}\sum_{J_\pi\;{\rm even}}
\left[\begin{array}{ccc}
j_\nu&j_\pi&J_1\\
j_\nu&j_\pi&J_2\\
J_\nu&J_\pi&J
\end{array}\right]
|J_\nu J_\pi;J\rangle,
\label{e_trans}
\end{equation}
where the symbol in square brackets is a unitary $9j$ coefficient~\cite{Talmi93}.
Note that the basis states on the right-hand side are orthogonal and normalized
while those on the left-hand side are not.
In the application of interest one constructs states in terms of a single np pair
such that $J_1=J_2\equiv\bar J$,
in which case the overlap matrix element is
\begin{equation}
\langle\!\langle\bar J^2;J|\bar J^2;J\rangle\!\rangle=
\frac{1+(-)^J}{4}\left(
1-\left[\begin{array}{ccc}
j_\nu&j_\pi&\bar J\\
j_\pi&j_\nu&\bar J\\
\bar J&\bar J&J
\end{array}\right]
\right),
\label{e_olap}
\end{equation}
which is zero for odd values of $J$.
This shows that the np pair behaves as a boson
since two identical np pairs can only coupled to even $J$.
Henceforth it is assumed that $J$ is even.
The matrix elements of the np interaction in the np--np basis are
\begin{eqnarray}
\langle\!\langle\bar J^2;J|\hat V_{\nu\pi}|\bar J^2;J\rangle\!\rangle&=&
V^{\bar J}_{\nu\pi}-
2\left[\begin{array}{ccc}
j_\nu&j_\pi&\bar J\\
j_\pi&j_\nu&\bar J\\
\bar J&\bar J&J
\end{array}\right]V^{\bar J}_{\nu\pi}
\nonumber\\&&+
(2\bar J+1)^2
\sum_R(2R+1)
\left[\begin{array}{cccccccc}
\!\!j_\nu\!\!&&\!\!j_\pi\!\!&&\!\bar J\!&&\!\bar J\!&\\
&\!\!R\!\!&&\!\!j_\nu\!\!&&\!J\!&&\!\!j_\pi\!\!\\
\!\!j_\nu\!\!&&\!\!j_\pi\!\!&&\!\bar J\!&&\!\bar J\!&
\end{array}\right]
V^R_{\nu\pi}.
\label{e_menpnp1}
\end{eqnarray}
For aligned np pairs, $J_{\rm max}=j_\nu+j_\pi$,
the unitary $9j$ coefficient in the overlap~(\ref{e_olap})
and in the matrix element~(\ref{e_menpnp1}) is much smaller than 1
and can be neglected.
[For $J=2J_{\rm max}$ the unitary $9j$ coefficient equals 1
but this angular momentum is excluded by the Pauli principle---see Eq.~(\ref{e_olap}).]
The sum over the $12j$ coefficients, however, cannot be neglected.
To a good approximation the matrix elements of the np interaction
between normalized states are therefore
\begin{equation}
\langle J_{\rm max}^2;J|\hat V_{\nu\pi}|J_{\rm max}^2;J\rangle\approx
2V^{J_{\rm max}}_{\nu\pi}+
2(2J_{\rm max}+1)^2
\sum_R(2R+1)
\left[\begin{array}{cccccccc}
\!\!j_\nu\!\!&&\!\!j_\pi\!\!&&\!J_{\rm max}\!&&\!J_{\rm max}\!&\\
&\!\!R\!\!&&\!\!j_\nu\!\!&&\!J\!&&\!\!j_\pi\!\!\\
\!\!j_\nu\!\!&&\!\!j_\pi\!\!&&\!J_{\rm max}\!&&\!J_{\rm max}\!&
\end{array}\right]
V^R_{\nu\pi}.
\label{e_menpnp}
\end{equation}
Not surprisingly, one finds that,
in a basis constructed out of aligned np pairs,
the main contribution to the 2n--2p configuration
stems from the matrix element $V^{J_{\rm max}}_{\nu\pi}$.
There are however corrections to this dominant contribution,
which depend on the total angular momentum $J$
and on the strengths of the np interaction with $R\neq J_{\rm max}$.

\section{Expressions in the high-$J$ and the low-$J$ limits}
\label{s_limits}
For a schematic short-range interaction the expressions in the np--np basis simplify.
Specifically, for a surface delta interaction (SDI)
classical approximations to the sum in Eq.~(\ref{e_menpnp})
can be worked out for aligned np pairs, $J_{\rm max}=j_\nu+j_\pi$.
With the method of Ref.~\cite{Isacker13},
based on the classical limit of $3nj$ coefficients~\cite{Wigner59},
the matrix element~(\ref{e_menpnp}) can be reduced to
\begin{equation}
\langle J_{\rm max}^2;J|\hat V_{\nu\pi}^{\rm SDI}|J_{\rm max}^2;J\rangle\approx
2V^{J_{\rm max}}_{\nu\pi}-\frac{3a_0+a_1}{\pi\sin\theta}+
(-)^{\ell_\nu+j_\nu+\ell_\pi+j_\pi}\frac{a_0-a_1}{\pi\tan\theta},
\label{e_meclas}
\end{equation}
where $a_0$ ($a_1$) is the isoscalar (isovector) strength of the np interaction
and $\theta$ is the angle between the angular momenta $J_{\rm max}$ of the two np pairs
that couple to total angular momentum $J$,
\begin{equation}
\theta=\arccos\frac{J(J+1)-2J_{\rm max}(J_{\rm max}+1)}{2J_{\rm max}(J_{\rm max}+1)}.
\label{e_theta}
\end{equation}

\begin{figure}
\centering
\includegraphics[width=0.3\textwidth]{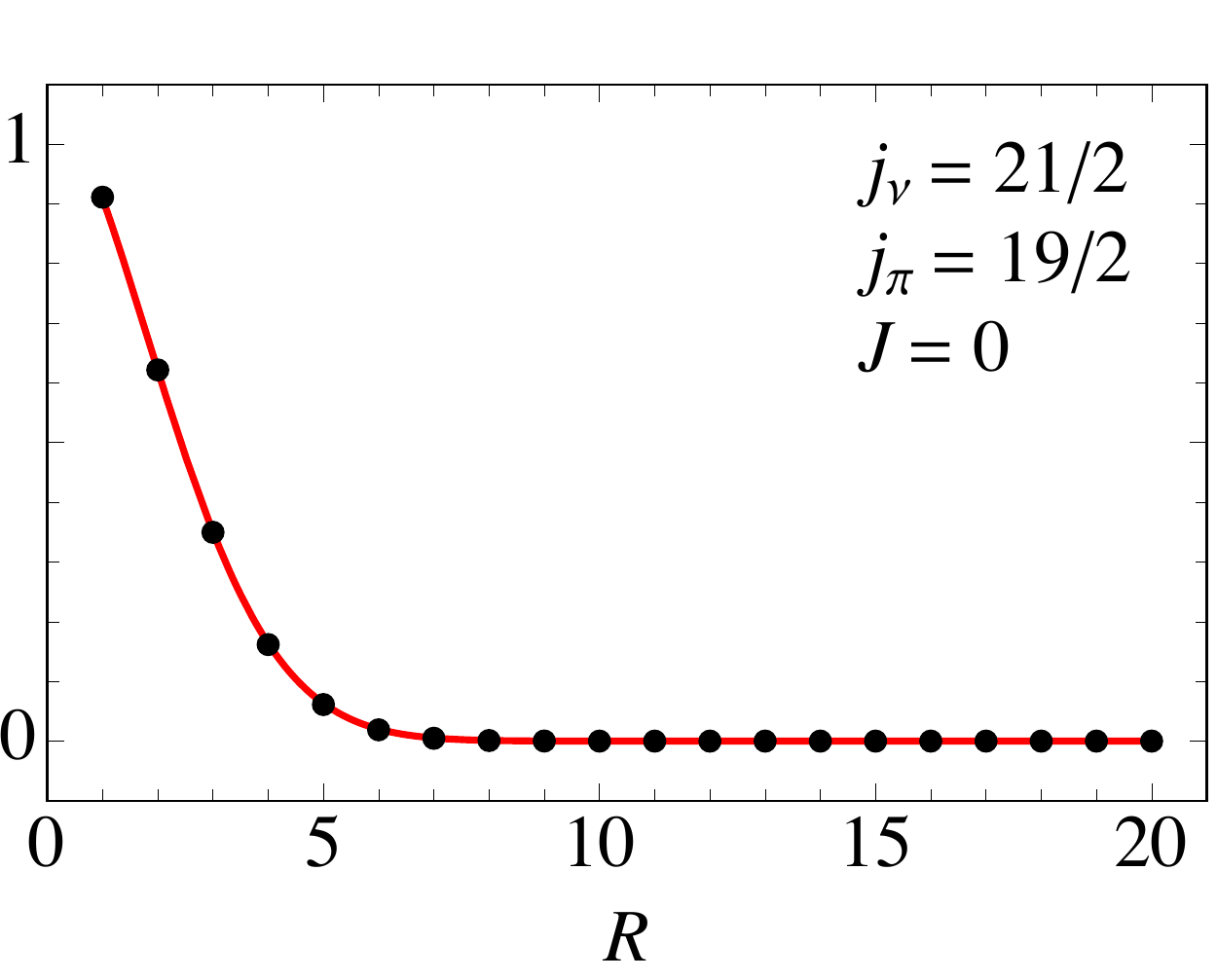} 
\includegraphics[width=0.3\textwidth]{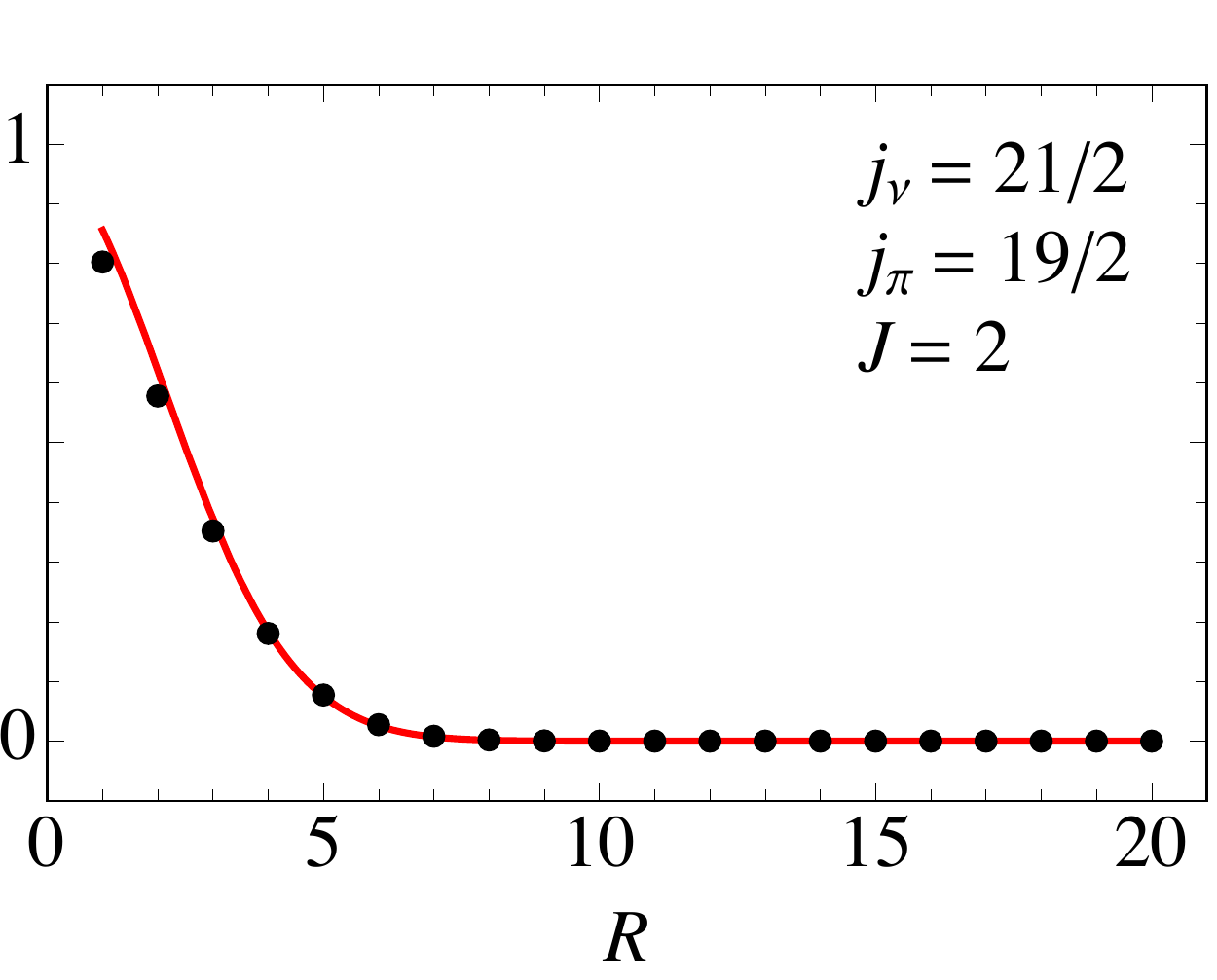} 
\includegraphics[width=0.3\textwidth]{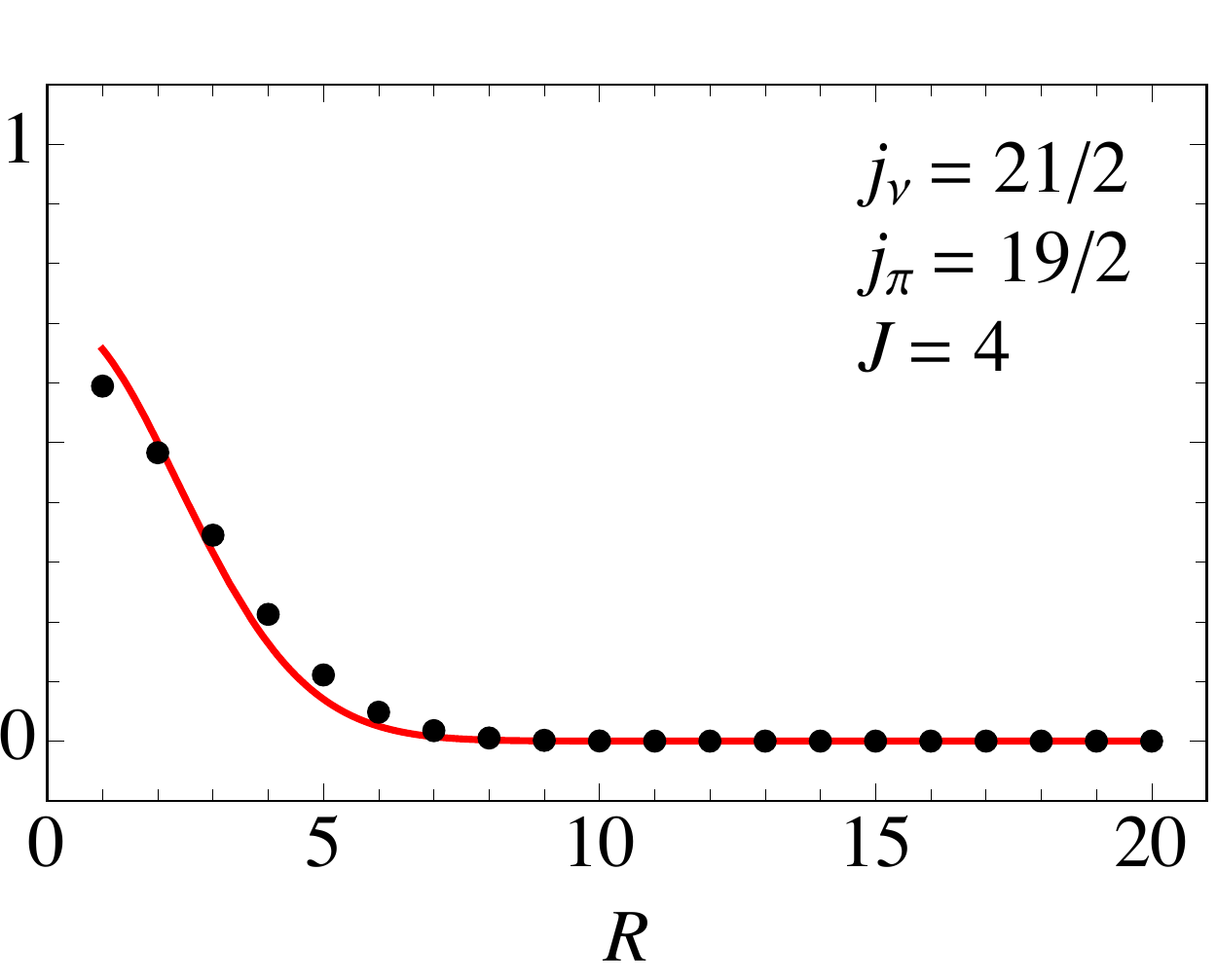}\\
\caption{Illustration of the $12j$-coefficient approximation
for $j_\nu=21/2$, $j_\pi=19/2$ and $J=0,2,4$, as a function of $R$.
The dots are the exact expression on the left-hand side
and the curve is its approximation on the right-hand side of Eq.~(\ref{e_12jnp}).}
\label{f_12j}
\end{figure}

\begin{figure}
\centering
\includegraphics[width=0.45\textwidth]{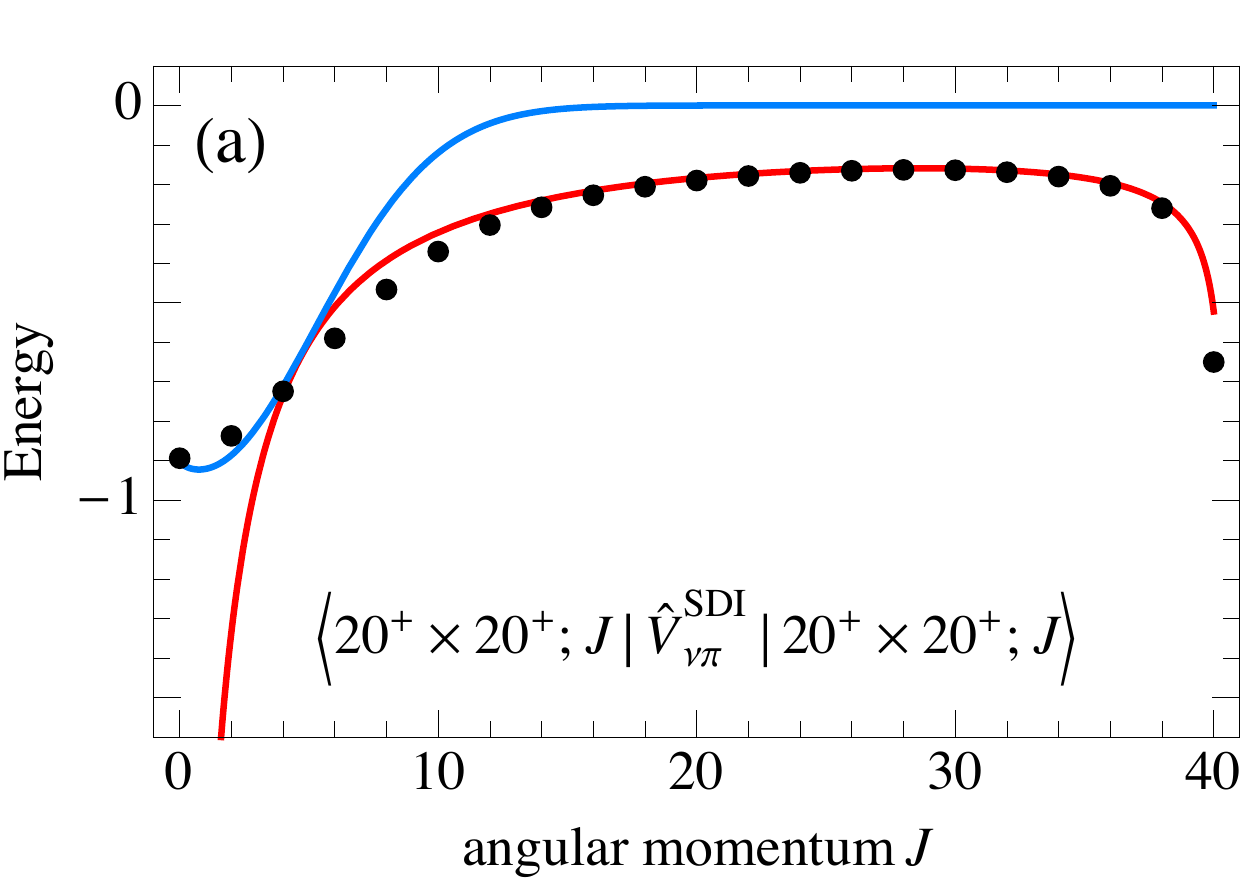} 
\includegraphics[width=0.45\textwidth]{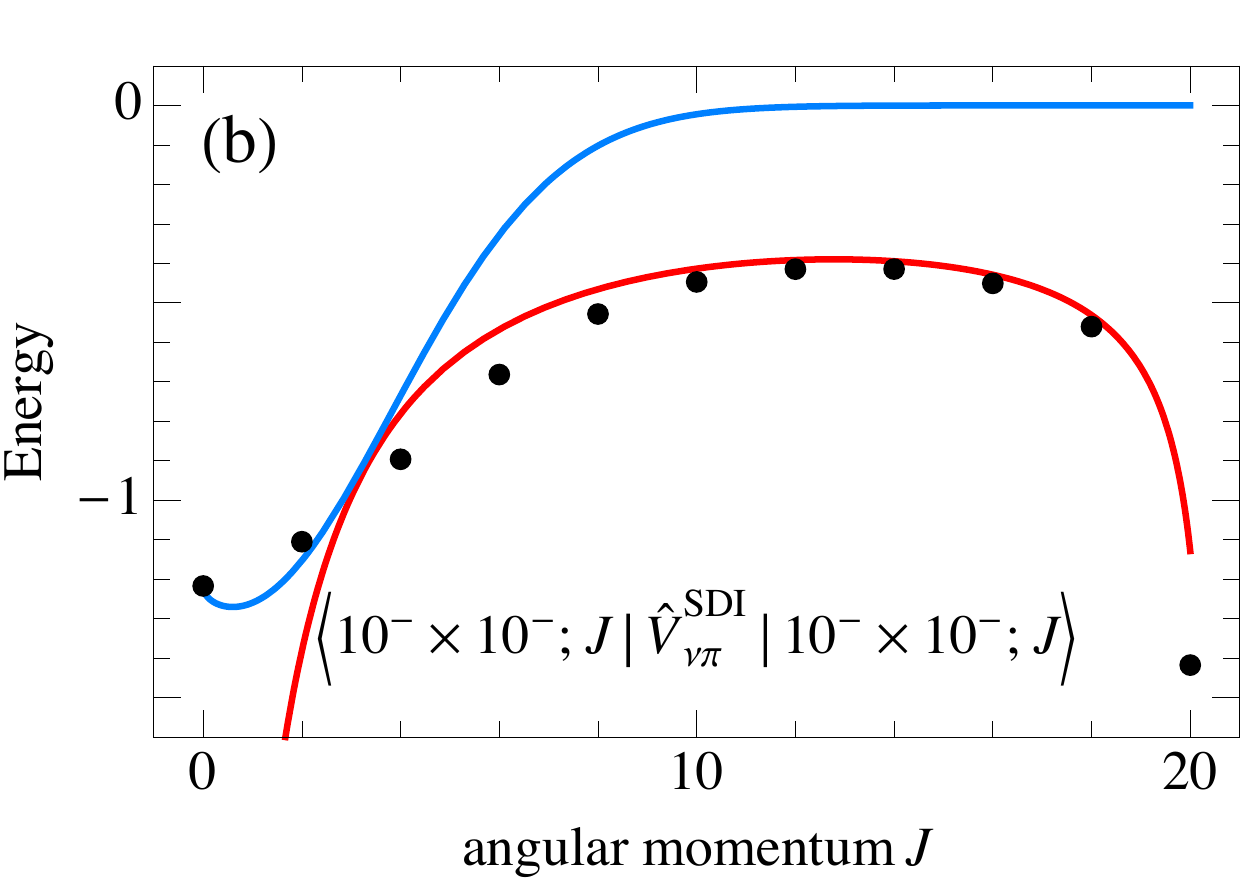} 
\caption{The exact expression for the 2n--2p matrix element~(\ref{e_menpnp}) of the np interaction (dots)
compared with the high-$J$ (red) and low-$J$ (blue) approximations
of Eqs.~(\ref{e_meclas}) and~(\ref{e_12jnp}), respectively.
The comparison is carried out for
(a) a SDI with $a_0=a_1=0.25$
and for $j_\nu=21/2$ and $j_\pi=19/2$ and $\ell_\nu$ and $\ell_\pi$ even,
and (b) a SDI with $a_0=0.75$ and $a_1=0.25$,
and for $j_\nu=h_{11/2}$ and $j_\pi=g_{9/2}$.
The constant contribution $2V_{\nu\pi}^{J_{\rm max}}$ is not included.}
\label{f_me21}
\end{figure}

The classical approximation~(\ref{e_meclas})
is reasonable except for low angular momentum $J$,
in particular for $J=0$ or $\theta=\pi$ when it diverges.
Unlike in Ref.~\cite{Isacker13} one is interested in the case $J=0$
since it corresponds to the energy of the ground state.
The following approximate expression
can be derived for the $12j$ coefficient
under the assumption that $J$ is low
and that the np pair is aligned, $J_{\rm max}=j_\nu+j_\pi$:
\begin{eqnarray}
\lefteqn{(J_{\rm max}+1)^2(2J_{\rm max}+1)
\left[\begin{array}{cccccccc}
\!\!j_\nu\!\!&&\!\!j_\pi\!\!&&\!J_{\rm max}\!&&\!J_{\rm max}\!&\\
&\!\!R\!\!&&\!\!j_\nu\!\!&&\!J\!&&\!\!j_\pi\!\!\\
\!\!j_\nu\!\!&&\!\!j_\pi\!\!&&\!J_{\rm max}\!&&\!J_{\rm max}\!&
\end{array}\right]}
\nonumber\\&&\qquad\approx
\exp\left[\frac{-4R(R+1)-J(J+1)+\sqrt{R(R+1)J(J+1)}+4(j_\nu-j_\pi)^2}{2(J_{\rm max}+1)}\right].
\label{e_12jnp}
\end{eqnarray}
This approximation is accurate for $J=0$
but deteriorates rapidly for increasing values of $J$,
unless the single-particle angular momenta are very large
(see Fig.~\ref{f_12j}).
The low-$J$ and high-$J$ approximations to the matrix element~(\ref{e_menpnp}) 
are illustrated in Fig.~\ref{f_me21}.

\section{The example of $^{128}$Cd}
\label{s_cd128}
\begin{figure}
\centering
\includegraphics[width=0.70\textwidth]{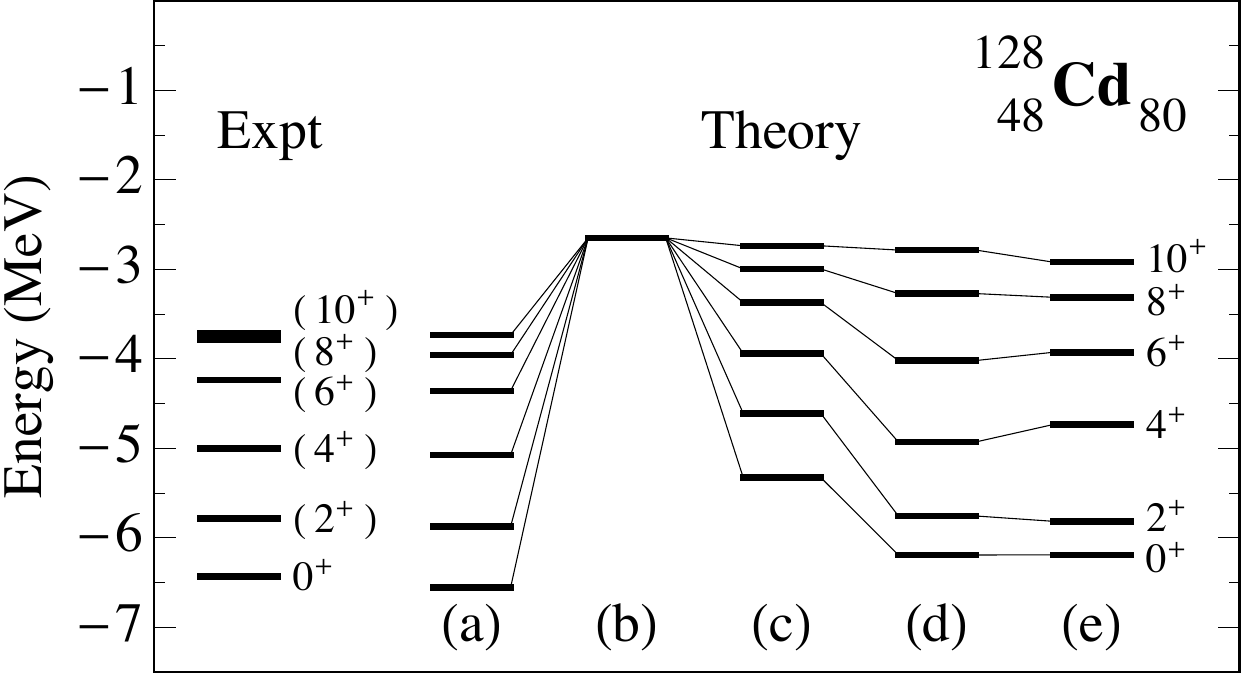} 
\caption{The experimental spectrum of $^{128}$Cd
compared to the shell-model approximations (a)--(e) discussed in the text.}
\label{f_cd128}
\end{figure}
Let me now illustrate the different approximations
with the example of the nucleus $^{128}$Cd,
with two neutron (proton) holes in $\nu0h_{11/2}$ ($\pi0g_{9/2}$)
with respect to the core $^{132}$Sn.
The relevant two-body np matrix elements
are derived from the realistic interaction jj45pna~\cite{Hjorthjensen95},
and the nn and pp interactions are taken from $^{130}$Sn and$^{130}$Cd, respectively.
The results of a shell-model calculation
with this interaction in the $(\nu0h_{11/2})^{-2}(\pi0g_{9/2})^{-2}$ model space
are shown in Fig.~\ref{f_cd128}(a),
and are seen to agree with the observed energies.
The jj45pna interaction can therefore be considered
as a realistic basis for further approximations.
The approximation (b) consists of keeping only the np matrix element $V^{10^-}_{\nu\pi}$
and putting all others (nn, pp and np) to zero.
As already pointed out by Zamick and Escuderos~\cite{Zamick13},
this clearly is inadequate.
It is essential to keep the nn and pp interactions
and therefore, in order to the test the influence of the np interaction,
in the approximations (c)--(e) the realistic nn and pp interactions
are taken without modification.
In Fig.~\ref{f_cd128}(c) are shown the results obtained
in the $(\nu0h_{11/2})^{-2}(\pi0g_{9/2})^{-2}$ model space
with the exact nn and pp interactions but with a single non-zero matrix element $V^{10^-}_{\nu\pi}$.
This is the approximation advocated in Ref.~\cite{Kimun};
it is indeed remarkable that with a {\em single} component of $V^R_{\nu\pi}$
that much of the np correlations can be accounted for.
The results (d) are obtained with Eq.~(\ref{e_menpnp})
and equivalent expressions for the nn and pp interactions (not given here).
This is the aligned-np-pair approximation~\cite{Isacker11}:
take the full, realistic interaction within the restricted space constructed out of aligned np pairs,
which in this example consists of a single state.
Comparison of (a) and (d) shows that the aligned-np-pair approximation
is adequate for low $J$ but fails for $J=8$ and 10.
Finally, the results (e) are obtained
with the np interaction in the low-$J$ approximation~(\ref{e_12jnp}) for $J\leq4$
and the classical approximation~(\ref{e_meclas}) for $J\geq4$ with $a_0=0.5$ and $a_1=0.13$,
strengths obtained from a fit of the SDI to the jj45pna interaction.
The results for $J=4$ are essentially indistinguishable in the two approximations.

\section{Concluding remark}
\label{s_conc}
It is important to distinguish between the following two approximations:
(i) The full shell-model interaction is considered in the restricted space constructed out of aligned np pairs;
(ii) Only the component of the np interaction in the aligned configuration is taken
and diagonalized in the full shell-model space.
The two approximations are {\em not} equivalent
and have their respective merits and problems.
If approximation (i) is made for a short-range interaction,
a simple geometric interpretation can be obtained,
as argued in this contribution.
Approximation (ii), on the other hand,
can be shown to give rise to a partially solvable problem
for the four-nucleon system~\cite{Isackerun}.

\section*{Acknowledgement}
I acknowledge fruitful discussions with
Yung Hee Kim, Maurycy Rejmund and Antoine Lemasson.


\begin{thebibliography}{99}
\bibitem{Schiffer71}
J.~P.~Schiffer,
Ann.\ Phys.\ (N.Y.) {\bf66} (1971) 798.

\bibitem{Molinari75}
A.~Molinari, M.~B.~Johnson, H.~A.~Bethe and W.~M.~Alberico,
Nucl.\ Phys.\ A {\bf239} (1975) 45.

\bibitem{Casten00}
R.~F.~Casten,
{\it Nuclear Structure from a Simple Perspective}
(Oxford University Press, Oxford, 2000).

\bibitem{Isacker14}
P.~Van~Isacker,
EPJ Web Conf.\ {\bf78} (2014) 03004.

\bibitem{Yutsis62}
A.~P.~Yutsis, I.~B.~Levinson and V.~V.~Vanagas,
{\it The Theory of Angular Momentum}
(Israel Program for Scientific Translations, Jerusalem, 1962).

\bibitem{Kimun}
Y.~H.~Kim, M.~Rejmund, P.~Van~Isacker and A.~Lemasson,
to be published.

\bibitem{Cederwall11}
B.~Cederwall {\it et al.},
Nature {\bf469} (2011) 68.

\bibitem{Frauendorf14}
S.~Frauendorf and A.~O.~Macchiavelli,
Prog.\ Part.\ Nucl.\ Phys.\ {\bf78} (2014) 24.

\bibitem{Isacker11}
P.~Van~Isacker,
Int.\ J.\ Mod.\ Phys.\ E {\bf20} (2011) 191.

\bibitem{Talmi93}
I.~Talmi,
{\it Simple Models of Complex Nuclei.
The Shell Model and the Interacting Boson Model}
(Harwood, Chur, 1993)

\bibitem{Wigner59}
E.~P.~Wigner,
{\it Group Theory and Its Application to the Quantum Mechanics of Atomic Spectra}
(Academic Press, New York, 1959).

\bibitem{Isacker13}
P.~Van~Isacker and A.~O.~Macchiavelli,
Phys.\ Rev.\ C {\bf87} (2013) 061301(R).

\bibitem{Hjorthjensen95}
M.~Hjorth-Jensen, T.~T.~Kuo and E.~Osnes,
Phys.\ Reports {\bf261} (1995) 125.

\bibitem{Zamick13}
L.~Zamick and A.~Escuderos,
Phys.\ Rev.\ C {\bf87} (2013) 044302.

\bibitem{Isackerun}
P.~Van~Isacker,
unpublished.

\end{thebibliography}
\end{document}